\pdfoutput=1

\documentclass[
 reprint,
 amsmath,
 amssymb,
 aps,
]{revtex4-1}
\usepackage{natbib}
\usepackage{graphicx}
\usepackage{dcolumn}
\usepackage{bm}
\usepackage{courier}
\usepackage{soul}
\usepackage{amsmath}
\usepackage{amssymb}
\usepackage{xcolor}
\usepackage[ruled, vlined]{algorithm2e}

\usepackage[utf8]{inputenc}

\begin{document}

\title{The Snake Optimizer for Learning Quantum Processor Control Parameters}

\author{Paul V. Klimov$^{1}$}\thanks{corresponding author, pklimov@google.com}
\author{Julian Kelly$^{1}$}
\author{John M. Martinis$^{2}$}
\author{Hartmut Neven$^{1}$}

\affiliation{
$^{1}$\text{Google AI Quantum}\\
$^{2}$\text{University of California, Santa Barbara}
}

\date{June 01, 2020}

\maketitle 

\textbf{High performance quantum computing requires a calibration system that learns optimal control parameters much faster than system drift. In some cases, the learning procedure requires solving complex optimization problems that are non-convex, high-dimensional, highly constrained, and have astronomical search spaces. Such problems pose an obstacle for scalability since traditional global optimizers are often too inefficient and slow for even small-scale processors comprising tens of qubits. In this whitepaper, we introduce the Snake Optimizer for efficiently and quickly solving such optimization problems by leveraging concepts in artificial intelligence, dynamic programming, and graph optimization.  In practice, the Snake has been applied to optimize the frequencies at which quantum logic gates are implemented in frequency-tunable superconducting qubits. This application enabled state-of-the-art system performance on a 53 qubit quantum processor, serving as a key component of demonstrating quantum supremacy. Furthermore, the Snake Optimizer scales favorably with qubit number and is amenable to both local re-optimization and parallelization, showing promise for optimizing much larger quantum processors.}

\section{\label{sec:Introduction}Introduction}
High-performance quantum processors require high-fidelity quantum logic gates.  In practice, logic gates are executed by manipulating the processor’s computing elements via choreographed control signals. In superconducting qubits, this typically amounts to sending shaped analog voltage and current pulses to manipulate qubits, couplers, and readout resonators \cite{QEG, Blais}. To achieve the precise system control necessary to execute high-fidelity gates, it is necessary to implement a calibration system that learns the circuit parameters of the computing elements and the optimal control signals that implement quantum logic, all much faster than system drift.

\begin{figure*}
\includegraphics[width=1\textwidth]{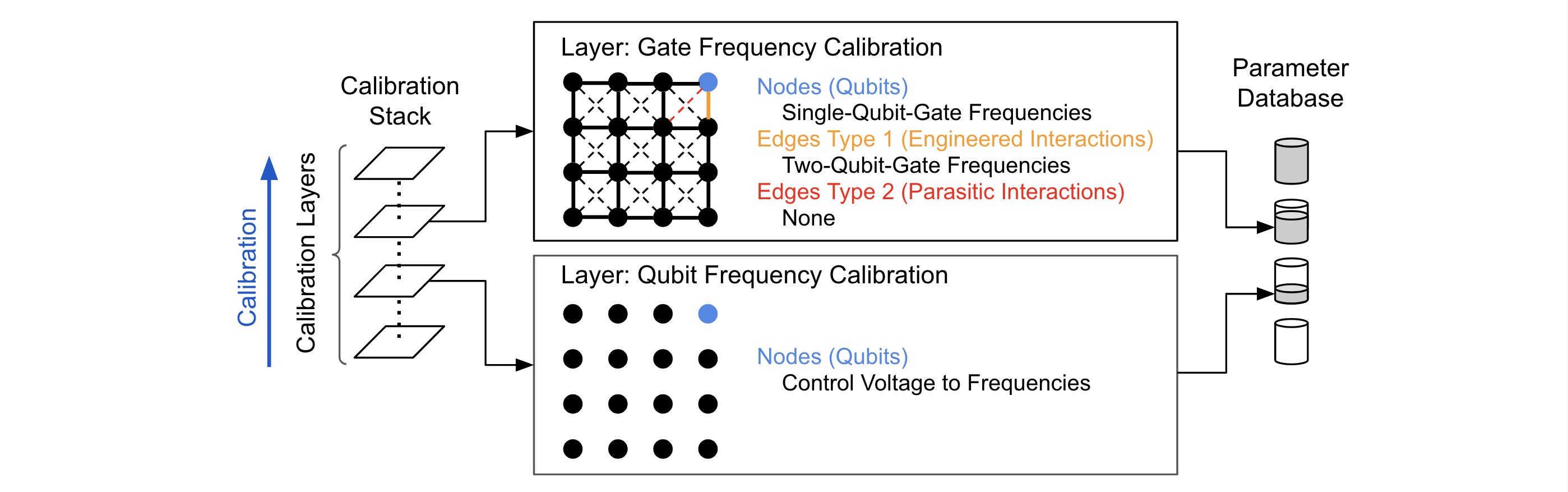}
\caption{\textbf{The Calibration System.} The calibration stack represents the calibrations that must be executed prior to quantum computation. At each layer of the stack, the calibration system learns the optimal parameters for some computing elements of the processor. The parameters are stored in a database for use when implementing quantum algorithms. At the start of calibration (lowest layer), the parameter database is empty. At the end of calibration (highest layer), the parameter database is full. An example independent calibration layer, where all optimal parameters are independent, is the qubit frequency calibration layer. Here, the calibration task is to learn the map from control electronics voltage to qubit frequency. An example interdependent calibration layer, where all optimal parameters depend on each other, is the gate-frequency calibration layer. Here, the calibration task is to choreograph single- and two-qubit gate frequencies over some quantum algorithm while mitigating computational errors. In practice, the Snake can be used to learn optimal parameters at arbitrary calibration layers.}
\label{fig: Introduction}
\end{figure*}

In practice, the calibration system is divided into three components (see Figure \ref{fig: Introduction}). The first component is the calibration stack, in which each layer represents one class of calibrations over the relevant computing elements. The second component is the system that learns the optimal parameters for the relevant computing elements at each layer. The third component is the system that navigates between calibration layers. The problem of constructing a calibration stack and navigating between it's layers has previously been addressed by Optimus \cite{optimus}. \textit{Here, we address the problem of learning optimal parameters at each calibration layer.}

How we learn optimal parameters at each calibration layer depends critically on the engineered and/or parasitic interactions between the relevant computing elements at that layer. When the interactions are negligible or can be ignored, the calibration layer is deemed \textit{independent} and each optimal parameter of each computing element can be learned independently and in parallel. An example independent calibration in frequency-tunable qubits is learning the map from control electronics voltages to qubit frequencies. Independent calibrations often map to low dimensional optimization problems that do not pose an obstacle for scalability. However, when the interactions are substantial and cannot be ignored, the calibration layer is deemed \textit{interdependent} and all optimal parameters of all computing elements depend on each other and must be learned simultaneously. Interdependent calibrations often map to complex combinatorial optimization problems that pose an obstacle for scalability, as illustrated next.

A key interdependent calibration layer is the quantum-logic gate frequency calibration layer. The calibration task in that layer is to choreograph all single and two-qubit gate frequencies over the course of the quantum algorithm for which the processor is being calibrated. The calibration is interdependent since all qubits can interact due to engineered interactions and/or parasitic crosstalk, and therefore, all optimal gate frequencies depend on each other either explicitly or implicitly. For a processor with $N$ nearest-neighbor coupled qubits on a square lattice, calibration thus maps to solving a combinatorial optimization problem over $O(N)$ dimensions in a $k^{O(N)}$ search space, where $k$ is the number of frequency options per gate. In practice $k \sim 10^2$, and so the search space significantly exceeds the processor's $2^N$ Hilbert space dimension. Given the problem complexity, exhaustive search is intractable and global optimization is inefficient for even small-scale processors with tens of qubits. We seek an efficient calibration strategy that can be applied to interdependent calibration layers.

We introduce the Snake Optimizer \cite{snake} as an efficient strategy for learning optimal parameters at arbitrary calibration layers.  For the nontrivial case of interdependent calibrations, the Snake can reduce one high-dimensional optimization problem into multiple lower-dimensional problems with exponentially reduced search spaces. The reduction in calibration complexity reduces the number of accessible solutions. Nonetheless, for many calibration layers, including gate-frequency calibration, we need only find one of many good solutions to operate a processor with state-of-the-art system performance.  

The Snake Optimizer was validated in Google's quantum supremacy demonstration with the 53 qubit Sycamore Processor \cite{QS}. When embedded into our calibration system, the Snake found a gate-frequency configuration for the full processor that outperformed a human expert by $\sim10^4\times$ in time and $\sim 20 \%$ in median two-qubit gate error, representing state-of-the-art system performance. Given its strong performance and favorable scaling in processor size, we are optimistic that the Snake will be a critical tool in optimizing and operating large-scale quantum processors.

\section{\label{sec:Snake}Snake Optimizer}
The Snake draws on ideas from artificial intelligence, graph optimization, and dynamic programming. It maps the parameters under calibration and the interactions between them onto the nodes and edges of a graph. Calibration is then accomplished by traversing the graph while calibrating a local subset of nodes and/or edges at each step. The step-by-step traversal across the processor is visually similar to the arcade game Snake, hence the name. Each calibration step is accomplished by building and optimizing an error model over the nodes and/or edges under calibration under constraints imposed by previously-calibrated nodes and/or edges, the hardware specifications, and the quantum algorithm for which the processor is being calibrated. This approach may be interpreted as greedy optimization that finds locally optimal solutions at each step. However, by applying dynamic constraints that depend on calibration history, the system emulates higher dimensional optimization, and returns a solution that satisfies all processor constraints. 

This whitepaper outlines a vanilla implementation of the Snake. In this implementation, we address the problem of finding optimal gate-frequencies for a superconducting quantum processor \cite{Koch, Barends} executing the cross-entropy benchmarking (XEB) quantum algorithm\cite{boixo, QS}. This implementation exercises the Snake machinery and the general ideas are extensible to any quantum processor architecture, calibration layer, and quantum algorithm. Moreover, they may be extensible to optimization problems beyond quantum computing, such as playing games, financial trading, or traffic routing.

\begin{figure}[t]
\includegraphics[width=\linewidth]{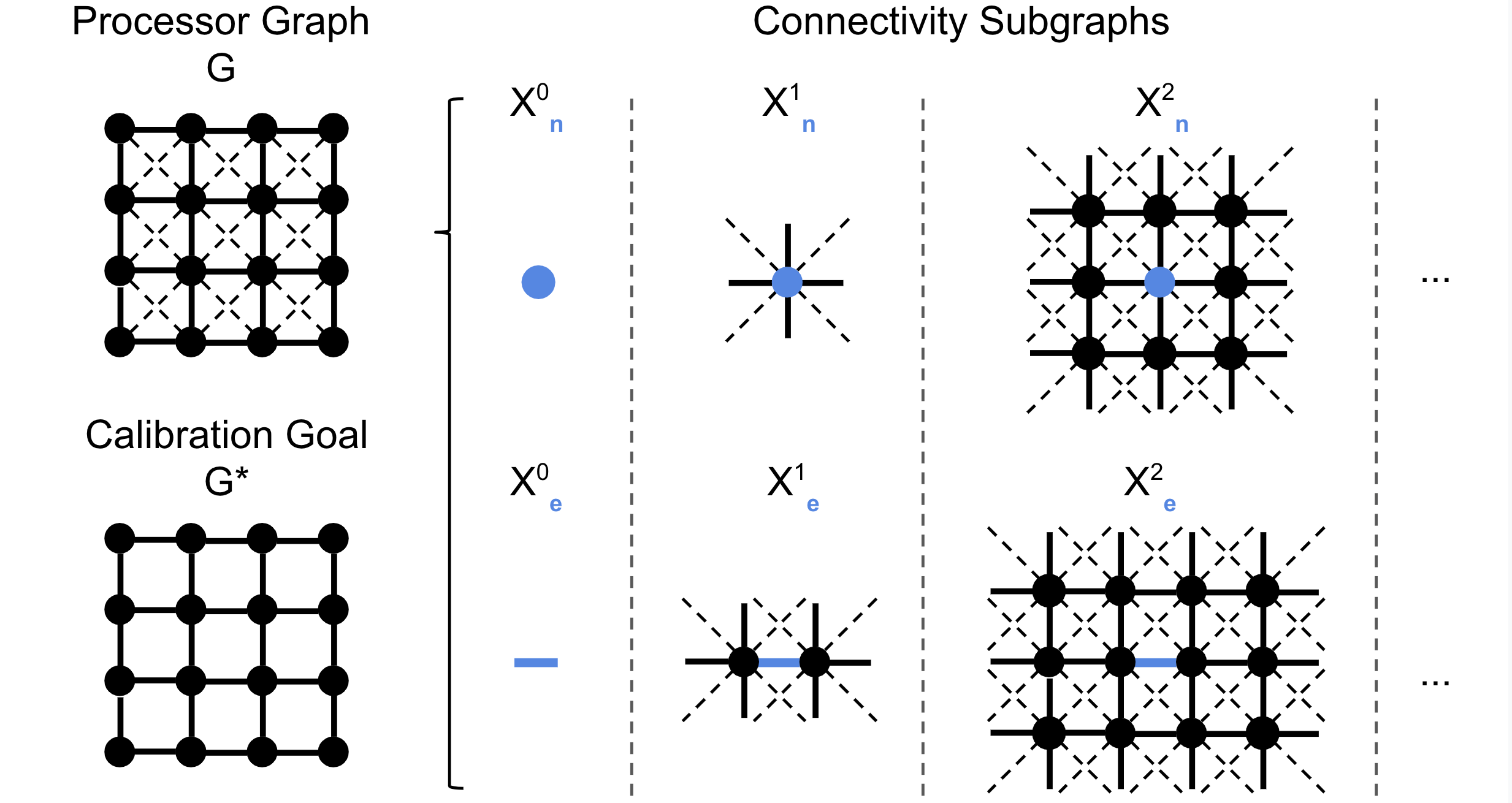}
\caption{\textbf{Processor Graph.} At each calibration layer, the relevant computing elements of the processor and the interactions between them are represented by the graph $G$. The graph elements that must be calibrated are represented by the calibration goal $G^*$. Here we show these structures for gate-frequency calibration. Nodes and horizontal/vertical edges have single- and two-qubit gate frequency trajectories associated with them that must be calibrated. Diagonal edges do not have parameters associated with them that must be calibrated, but are included in $G$ to embed crosstalk constraints. Each graph element's connectivity is defined via a connectivity subgraph $X^d_g$, where $d$ is an edge-distance that specifies scope. Connectivity subgraphs are used to drive graph traversal and build calibration parameters and constraints.}
\label{fig: Graph}
\end{figure}

\subsection{\label{sec:Graph Representation}Processor Graph}
At each calibration layer, we map the relevant computing elements and interactions between them onto the nodes $N$ and edges $E$ of an undirected \textit{graph} $G = N \cup E$ (Figure \ref{fig: Graph}). Nodes $n\in N \subseteq G$ typically represent qubits and edges $e\in E \subseteq G$ typically represent the interactions between them. Different edge types are used to distinguish engineered interactions from parasitic crosstalk. Arbitrary nodes or edges are referred to as graph elements $g$. The connectivity of any $g$ is defined by a \textit{connectivity subgraph} $X^{d}_g \subseteq G$, which we define as the set of graph elements within $d$ edge-traversals of $g$ (Figure \ref{fig: Graph}). The parameter $d$ is thus a distance that specifies scope.

The purpose of calibration is to learn the optimal parameters for some subset of nodes and/or edges, which we refer to as the \textit{calibration goal} $G^*\subseteq G$. Calibration is accomplished by traversing $G^*$ while calibrating the parameters associated with some subset of nodes and/or edges $P\subseteq G^*$ at each step under respective calibration constraints $R_P \subseteq G^*$. The parameters that are calibrated at each step are added to the \textit{calibration status} $P^*\subseteq G^*$, which is an ordered set that encodes calibration history. Un-calibrated graph elements are $G^* \setminus P^*$. Calibration is complete when $P^*= G^*$. 

For gate-frequency calibration, the node parameters are single-qubit gate frequencies and the engineered edge parameters are two-qubit gate frequency trajectories. Parasitic edges are not calibrated but are included in $G$ to inform constraints. Calibration is complete when all optimal gate frequencies have been learned. Prescriptions for driving graph traversal and building calibration structures are presented in forthcoming sections. 

\subsection{\label{sec:Algorithm Subgraphs}Algorithm Subgraphs}

\begin{figure}[b]
\includegraphics[width=\linewidth]{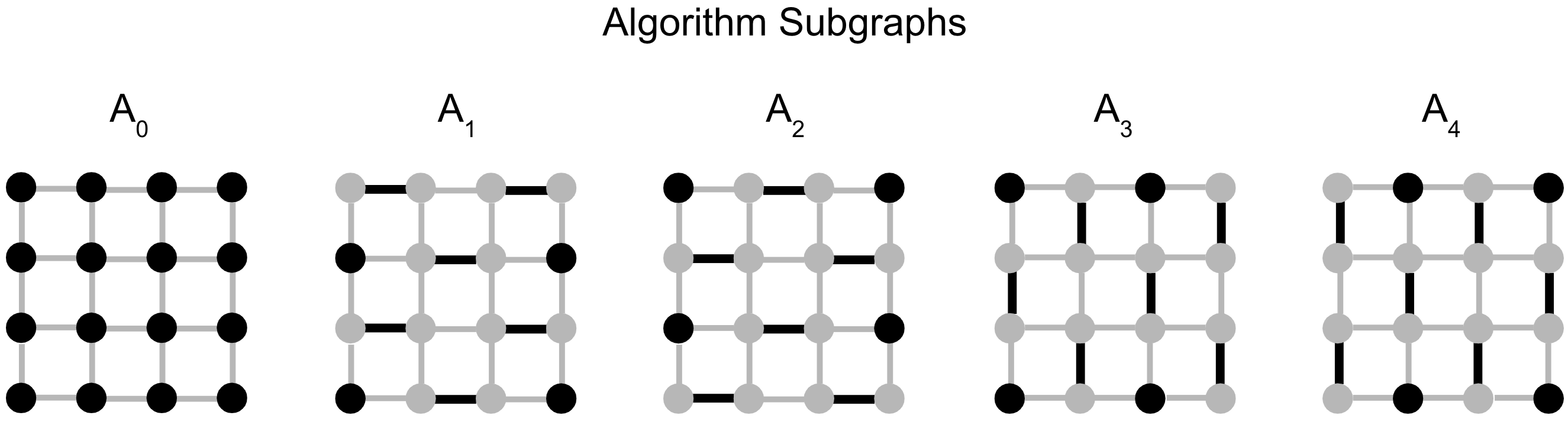}
\caption{\textbf{Algorithm Subgraphs}. Algorithm subgraphs $A_{i}$ represent the simultaneously active graph elements at distinct temporal moments of the quantum algorithm for which the processor is being calibrated. They are used to determine which graph elements can be active at the same time as the relevant calibration parameters and thus apply constraints. Here we show the algorithm subgraphs for XEB in the context of gate frequency calibration. XEB contains alternating layers of single- and two-qubit gates up to arbitrary depth. Even though gate parameters, such as rotation angles and phases, can differ between temporal moments, their frequency configurations are constant, leading to 5 distinct subgraphs.}
\label{fig:XEB_Algorithm_Subgraphs}
\end{figure}

We embed the spacetime structure of the quantum algorithm for which the processor is being calibrated via a set of \textit{algorithm subgraphs} $\{A_i | A_i \subseteq G^*\}$. Each subgraph comprises the simultaneously active graph elements at a distinct temporal moment of the quantum algorithm. In other words, they contain the elements that can potentially interfere and thus constrain each other. What the algorithm subgraphs represent physically depends on the processor architecture and the calibration layer. 

For gate-frequency calibration, the algorithm subgraphs represent the logic gates that can be executed simultaneously. For fully unstructured quantum computation, most graph elements can be active simultaneously and thus constrain each other. This case may be treated approximately with a single algorithm subgraph $A_0 = G^*$. However, in algorithms with spacetime symmetry such as XEB, only some graph elements can be active simultaneously and thus constrain each other. This case may be treated with the algorithm subgraphs presented in Figure \ref{fig:XEB_Algorithm_Subgraphs}. We exploit such symmetries to learn better parameters, as described next. 

At each calibration step, we build the \textit{active graph elements} $A_{g} = \bigcup_i (A_i|g\in A_i)$ for each element of the calibration parameters $g \in P$. The active elements for each $g$ are then used to filter the calibration constraints down to those that can be simultaneously active as $g$ (see Section  \ref{sec:Calibration_Constraints}). This filtration procedure results in more physical constraints and thus better learned parameters. 

\subsection{\label{sec:Graph Traversal}Graph Traversal}
To calibrate a graph, we traverse it while calibrating graph elements at each step. However, since ${G}$ is undirected, there is no preferred traversal path. We implicitly direct $G$ via a \textit{traversal rule} and \textit{traversal heuristic}.  The traversal rule takes the \textit{central graph element} at each traversal step and returns a set of candidate graph elements to traverse next. The traversal heuristic then sorts those elements, typically based on traversal history, to drive the desired traversal, for example breadth-first, depth-first, or random. The sorted graph elements are referred to as \textit{traversal options}. 

In practice, we build traversal options for arbitrary central graph element $g$ via its $d_T$ connectivity subgraph as $\text{heuristic}(g' | g' \in X_g^{d_T} \cap A_{g} \cap G^* \setminus P^* \text{ and } \text{type}(g') = \text{type}(g) \text{ and } g \leftrightarrow g')$. The first condition drives traversal towards un-calibrated graph elements that can be simultaneously active. The second condition drives traversal towards graph elements of the same type. The third conditions ensures that traversal is \textit{symmetric}, such that if $g$ can traverse to $g'$, then $g'$ can traverse to $g$. Traversal options are built via some function \texttt{build\_traversal\_options}.

\begin{figure}[t]
\includegraphics[width=\linewidth]{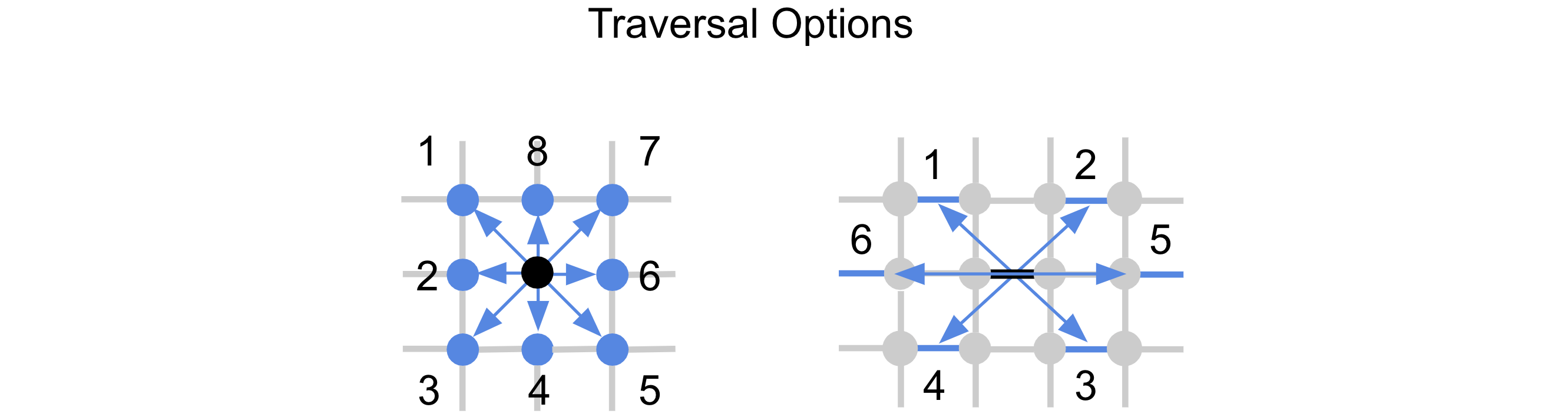}
\caption{\textbf{Graph Traversal.} The traversal options drawn at each step are defined via a traversal rule and traversal heuristic. Together, these are chosen to implement the desired traversal behavior, such as breadth-first, depth-first, or random. Here we show the traversal options for an arbitrary central node and edge with $d_T=2$ for XEB. The numbers represent the sorting order chosen via an arbitrary heuristic.}
\label{fig:Traversal_Options}
\end{figure}

\subsection{\label{sec:Traversal Threads}Graph Segmentation and Seeding}

Having implicitly directed $G$ for traversal, we now segment the un-calibrated elements $G^* \setminus P^*$ two times to generate useful calibration structures that offer a route to parallelization and ensure complete traversal. First, we segment the un-calibrated elements into \textit{calibration subgoals}, which can be interpreted as sufficiently distant regions of the processor that can be treated independently and calibrated in parallel. Second, we segment each calibration subgoal into \textit{traversal threads}, which are isolated from the perspective of graph traversal, under the chosen traversal rule. Each thread in each subgoal must be seeded and traversed to completion once and only once to traverse all un-calibrated elements once and only once. Below we describe how these structures are built.

\begin{figure}[b]
\includegraphics[width=\linewidth]{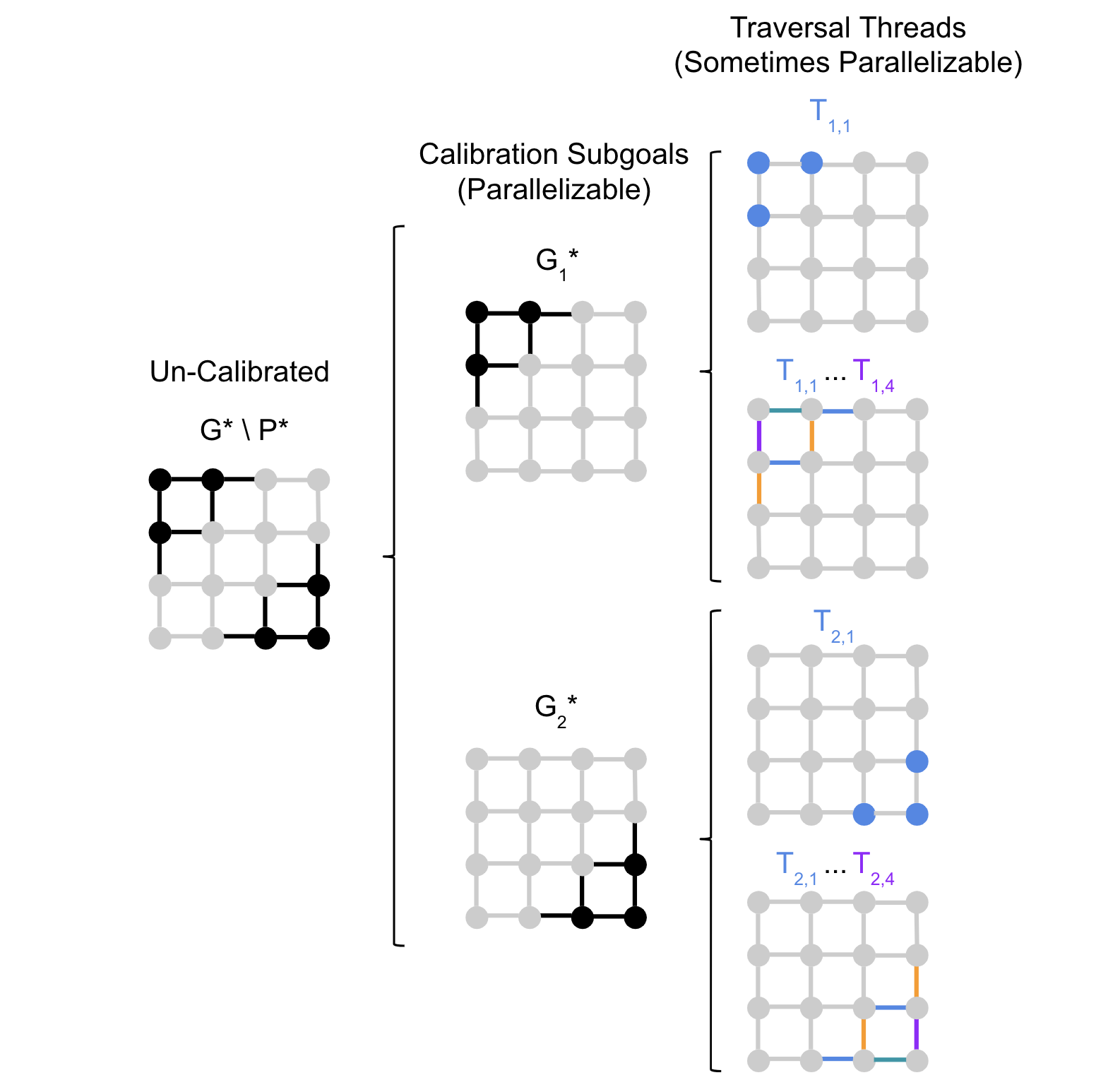}
\caption{\textbf{Graph Segmentation and Seeding}. Prior to calibration, we determine which regions of a processor are isolated enough to be calibrated in parallel and how to traverse all elements of those regions. We do so by segmenting the un-calibrated elements $G^* \setminus P^*$ into calibration subgoals $G_i^*$ and respective traversal threads $T_{i,j}$. Here we segment a processor with an arbitrarily chosen calibration status under the traversal rule described in Figure \ref{fig:Traversal_Options}. The segmentation results in 2 calibration subgoals, which can be calibrated in parallel. Each subgoal comprises 1 node traversal thread and 4 edge traversal threads, which are color coded. Each thread must be seeded and traversed to completion to traverse each un-calibrated element. The traversal threads may be calibrated in parallel depending on factors sketched in the text.}
\label{fig:Disjoint_Tree}
\end{figure}

The calibration subgoals $G_i^*$ are built to satisfy $\{G_i^* \subseteq G^* \setminus P^* | G^* \setminus P^* = \bigcup_i G^*_i \text{  and } R_{G^*_i} \cap G^*_j = \emptyset \forall i \neq j\}$. The first condition ensures that the combined calibration subgoals comprise all un-calibrated elements of the calibration goal. The second condition ensures that the calibration subgoals are \textit{constraint disjoint}, which means that all constraints of all elements in a given subgoal are disjoint from all elements of any other subgoal. This condition thus ensures that each calibration subgoal can be calibrated in parallel without interference. An important subtlety is that calibration subgoals may share previously calibrated graph elements and thus constraints. Formally this means that the cardinality $|R_{G^*_i} \cap R_{G^*_j}|\geq0$. We build calibration subgoals via some function \texttt{build\_calibration\_subgoals}.

The traversal threads $T_{i,j}$ for calibration subgoal $G_i^*$ are generated by the chosen traversal rule to satisfy $\{T_{i,j} \in G^*_i | G^*_i = \bigcup_j T_{i,j} \text{  and  } T_{i,j} \cap T_{i',j'} = \emptyset \forall (i,j)\neq (i',j')\}$. The first condition ensures that the combined traversal threads in a calibration subgoal comprise all elements of that subgoal. The second condition ensures that threads are \textit{traversal disjoint}, which means that the traversal rule cannot drive traversal between them. In combination with the requirement that traversal options must be drawn symmetrically, this condition further implies that each thread must be seeded once and only once to traverse all un-calibrated elements once and only once. An important note is that for some architectures, calibration layers, quantum algorithms, and calibration and traversal parameters, the traversal threads in a given subgoal may also be constraint-disjoint. In that case, their calibration may also be parallelized.  We build traversal threads via some function \texttt{build\_traversal\_threads} and pick a seed from each thread via some function \texttt{build\_traversal\_seed}.

\begin{figure*}
\includegraphics[width=\textwidth]{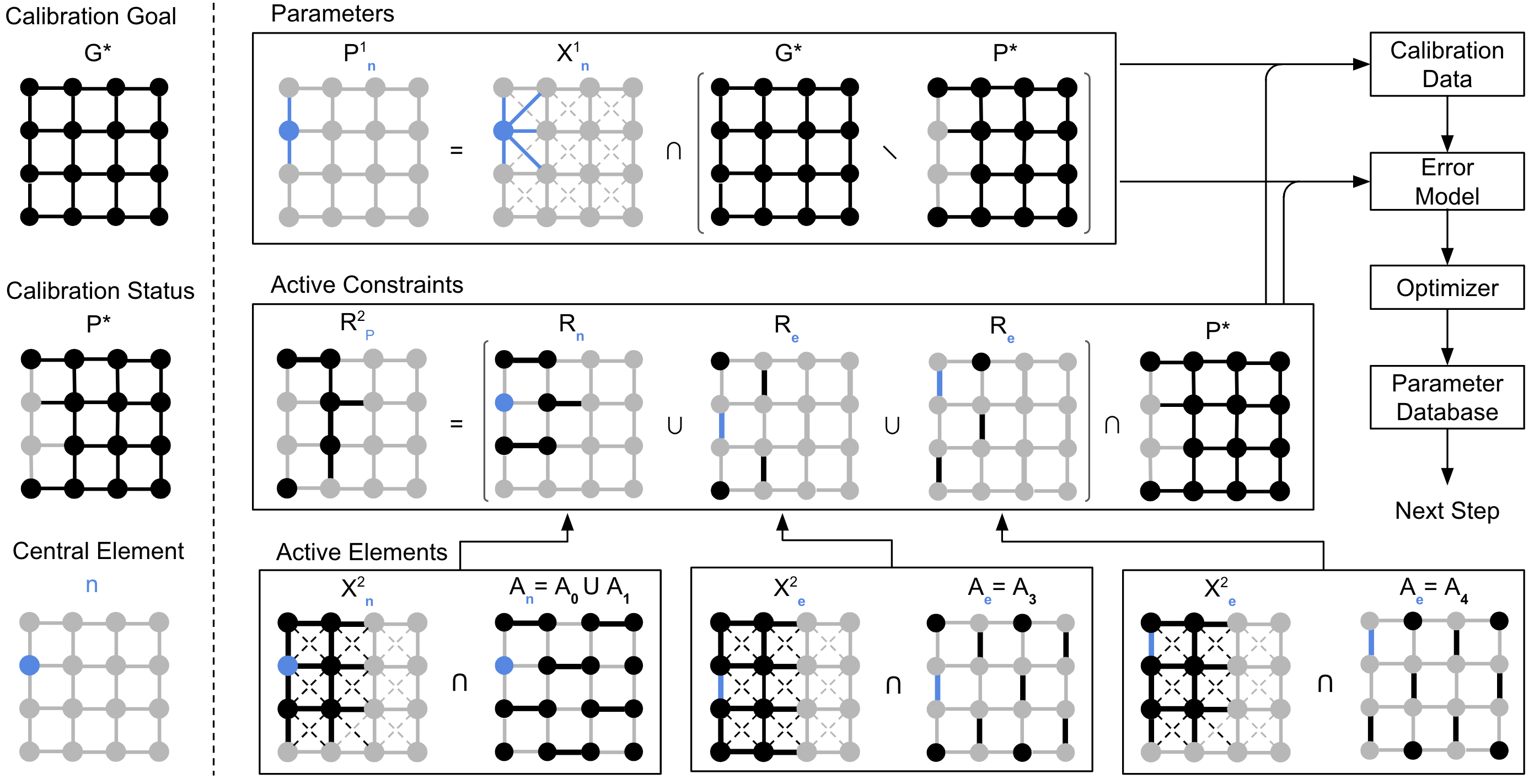}
\caption{\textbf{Calibration Structures at One Calibration Step}. At each traversal step, we build calibration parameters and active calibration constraints around the central graph element. These structures directly inform which calibration data are relevant and how to build an error model around that data. The error model is then optimized with respect to the calibration parameters to learn optimal control parameters. Those parameters are stored in the parameter database and calibration proceeds onto the next step. Here we show the calibration structures for an arbitrarily chosen calibration goal $G^*$, calibration status $P^*$, and central graph element $n$. (upper) Calibration parameters $P^1_n$ built around $n$ with scope $d_P=1$. For this configuration, 1 node and 2 edges are calibrated simultaneously, which corresponds to 3-dimensional optimization. (central) Active calibration constraints $R^2_{P}$ with scope $d_R=2$ for XEB. These constraints are constructed from the active graph elements as shown in the (lower) panel. For this configuration, the calibration parameters are constrained by 5 nodes and 5 edges. Similar structures are built at each traversal step of the calibration procedure.}
\label{fig:Calibration Structures}
\end{figure*}

\subsection{\label{sec:Calibration Parameters}Calibration Parameters}
At each traversal step, we build a set of \textit{calibration parameters}, which are the graph elements whose optimal parameters are learned and thus calibrated at that step. We build calibration parameters from the central graph element $g$ via it's distance-$d_P$ connectivity subgraph as $P^{d_P}_g = X^{d_P}_g \cap G^* \setminus P^*$ (see Figure \ref{fig:Calibration Structures}). The intersection ensures that previously calibrated graph elements are not re-calibrated. We build calibration parameters via some function \texttt{build\_calibration\_parameters}. 

A key parameter of the Snake is the distance $d_P$, which controls optimization scope and thus complexity. Namely, the number of calibration parameters $|P^{d_P}_g|$ at any step ranges between 1 and $\sim d_P^2$, depending on graph structure and calibration history. Therefore, full calibration requires between $|G^* \setminus P^*|$ and $\sim |G^* \setminus P^*|/d_P^2$ steps, each of which corresponds to $|P^{d_P}_g|$-dimensional optimization over search-space $k^{|P^{d_P}_g|}$. Here $k$ is the number of options for each $g \in P^{d_P}_g$, which is assumed independent of $g$ for simplicity. To understand the implications of $d_P$ on calibration complexity, we consider two limits. 

In the limit when $d_P \sim \sqrt{N}$, calibration takes 1 step, with all graph elements $P^{d_P}_g = G^* \setminus P^*$ calibrated simultaneously. Calibration thus corresponds to $|G^* \setminus P^*|$-dimensional optimization over search-space $k^{|G^* \setminus P^*|}$. This is the full global optimization problem that the Snake was developed to simplify, which we do by reducing $d_P$. 

In the limit when $d_P=0$, calibration takes $|G^* \setminus P^*|$ steps, with only the central graph element $P^{0}_g = g$ calibrated at each step. Each problem thus corresponds to $1$-dimensional optimization over search-space $k$. Therefore, this limit offers a massive reduction in complexity over global optimization. The trade-off is in the number of accessible solutions, which is suppressed exponentially.

The Snake can therefore interpolate between optimization complexity, establishing itself as a flexible and powerful calibration strategy. The trade-off between optimization complexity and the number of accessible solutions is significant and should be made based on factors including the quantum processor architecture, the calibration layer, the quantum algorithm for which the processor is being calibrated, it's performance requirements, and access to classical compute resources. 

\subsection{\label{sec:Calibration_Constraints}Calibration Constraints}
At each traversal step, we build a set of constraints on the calibration parameters. There is flexibility in the constraints' physical origin, their scope, and their mathematical representation in the error model. For gate-frequency calibration, the constraints may originate from control or qubit-qubit crosstalk, the control hardware specifications, or qubit circuit parameters. Their scope may comprise only the calibration parameters or all elements within several edge traversals of those parameters. Finally, the constraints may be represented as smooth functions or hard bounds.  In any case, the constraints must be built such that the full calibration procedure returns learned parameters that respect all processor constraints simultaneously, regardless of calibration order. 

We build the \textit{active calibration constraints} for an arbitrary parameter set $P$ as $R_P^{d_R} = P^* \cap \bigcup_{g\in P} X_g^{d_R} \cap A_g$ (see Figure \ref{fig:Calibration Structures}). The intersection with $P^*$ ensures that only calibrated graph elements apply constraints. Intuitively, the un-calibrated elements cannot apply constraints since their parameters have not yet been learned. The intersection with $A_g$ ensures that only simultaneously active graph elements apply constraints. A final and critical requirement is that constraints are symmetric, such that if $g$ constrains $g'$, then $g'$ constrains $g$. We build active calibration constraints via some function \texttt{build\_constraints}.

\subsection{\label{sec:Error Model}Calibration Error Model}
At each traversal step, we build an error model that maps the calibration parameters onto some system performance metric. That error model is constructed from the calibration parameters, the active calibration constraints, and the relevant calibration data. The physical error mechanisms encompassed by the model and their mathematical representation are typically determined through physics research and machine learning. Moreover, they depend strongly on the processor architecture, the calibration layer, and the quantum-algorithm for which the processor is being calibrated. For gate-frequency calibration in superconducting qubits, the model may capture frequency and quantum algorithm dependent relaxation, dephasing, leakage, and control errors \cite{QS}. We build calibration data via some function \texttt{build\_calibration\_data}. We build the error model via some function \texttt{build\_error\_model}. Finally, we learn optimal calibrated parameters via an arbitrary optimization subroutine via some function \texttt{optimize\_error\_model}.

\begin{algorithm}[t]
\textbf{args} 
$
\begin{cases}
G - \text{Processor Graph}\\
G^{*} - \text{Calibration Goal}\\
P^{*} - \text{Calibration Status}\\
d_{P} - \text{Parameter Distance}\\
d_{T} - \text{Traversal Distance}\\
d_{R} - \text{Constraint Distance}\\
\text{Traversal Rule}\\
\text{Traversal Heuristic}\\
\text{Quantum Algorithm}\\
\text{Optimizer}\\
\end{cases}
$

\texttt{\\}

\textbf{def} \text{calibrate\_element}($g$, \textbf{*args}):\\
\quad $P_{g}^{d_P}$ = \text{build\_parameters($g$, \textbf{*args})}\\
\quad $R_{P}^{d_R}$ = \text{build\_constraints}(\textbf{$P_{g}^{d_{P}}$, *args})\\
\quad $D$ = \text{build\_calibration\_data($P_g^{d_P}$, $R_{P}^{d_R}$, \textbf{*args})}\\
\quad \text{model} = \text{build\_error\_model($P_g^{d_P}$, $R_{P}^{d_R}$, $D$, \textbf{*args})}\\
\quad \text{optimize\_error\_model(model, \textbf{*args})}\\
\quad $P^* = P^* \cup P_g^{d_P}$

\texttt{\\}
\textbf{def} calibrate\_thread($g$, \textbf{*args}):\\
\quad \textbf{if} $g \notin P^*$: \\
\quad \quad \text{calibrate\_element}($g$, \textbf{*args})\\
\quad \quad \textbf{for} $g' \in$ \text{build\_traversal\_options}(g, \textbf{*args}):  \\
\quad \quad \quad    calibrate\_thread($g'$, \textbf{*args})\\
\texttt{\\}
\textbf{def} calibrate\_graph(\textbf{*args}):\\
\quad \textbf{parfor} $G_i^* \in$ \text{build\_calibration\_subgoals(\textbf{*args})}:

\quad \quad \textbf{for} $T_{i,j} \in$  \text{build\_traversal\_threads($G_i^*$, \textbf{*args})}:

\quad \quad \quad $g$ = \text{build\_traversal\_seed}($T_{i,j}$, \textbf{*args})\\
\quad \quad \quad \text{calibrate\_thread}($g$, \textbf{*args})\\
\caption{Snake Optimizer}
\label{alg:Graph_Calibration}
\end{algorithm}

\begin{figure}[b!]
\includegraphics[width=\linewidth]{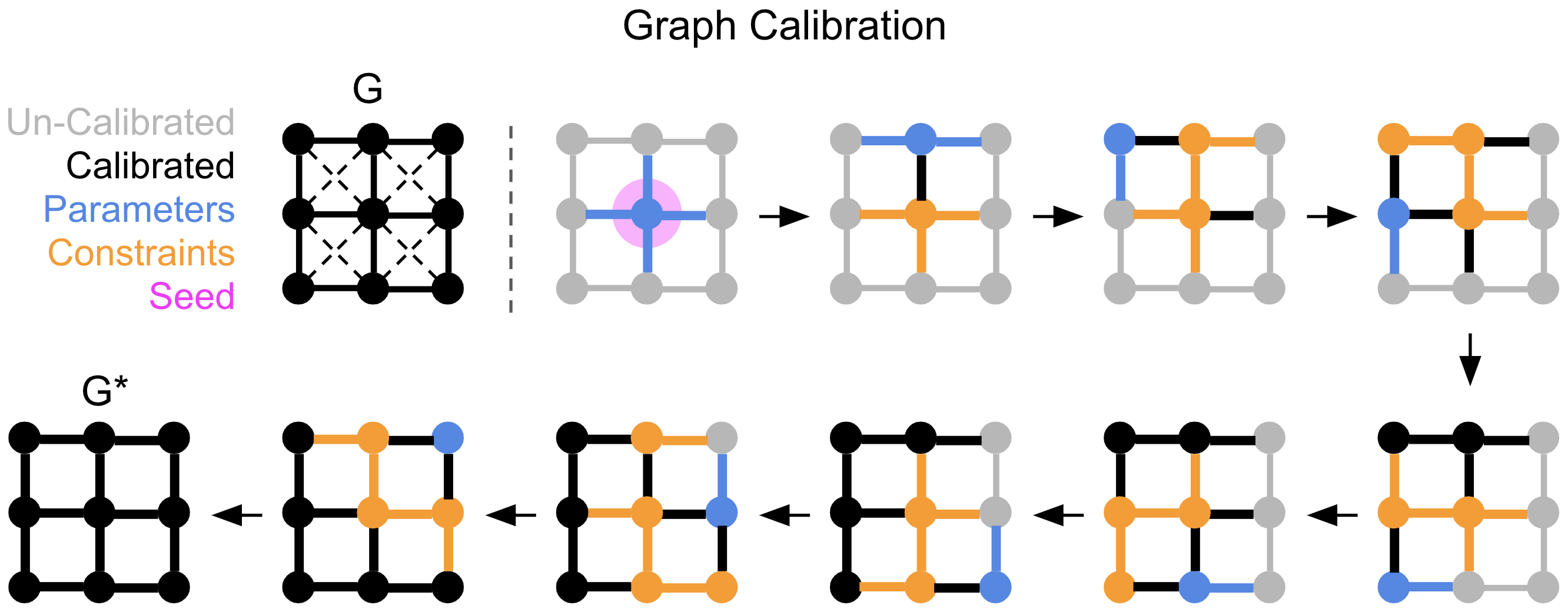}
\caption{\textbf{Graph Calibration}. The Snake applied to optimize graph $G$ at the gate frequency calibration layer for the XEB quantum algorithm. Un-calibrated elements are grey, calibrated elements are black, calibration parameters are blue, and active calibration constraints are orange. The parameter, constraint, and traversal distances are set to $d_P=1$, $d_R=2$, $d_T=2$, respectively. The graph is seeded on the pink node and traversal is driven via an arbitrarily chosen traversal rule and heuristic. Even though there are multiple traversal subgraphs, 1 node seed is sufficient for complete calibration due to the chosen traversal rule and calibration parameter distance. Over the course of the calibration procedure, the optimization dimension ranges from 1 to 5 and the number of active constraints ranges from 0 to 8, depending on the central graph element and the calibration history.}
\label{fig:Node then Edge Calibration}
\end{figure}

\subsection{\label{sec:Graph Calibration}Graph Calibration}
We now consolidate the concepts introduced above to define a vanilla implementation of the Snake Optimizer (see Algorithm \ref{alg:Graph_Calibration}). The strategy can be summarized as follows: First we segment the calibration goal into constraint-disjoint calibration subgoals, which launches isolated and parallelizable calibration regions. We then split each calibration subgoal into traversal-disjoint traversal threads, which we seed and recursively traverse while calibrating graph elements at each step. Note that an iterative approach can be used with some trade-offs. This strategy is illustrated in Figure \ref{fig:Node then Edge Calibration}.

This vanilla implementation leaves much room for advancements.  One extension is to cache error model evaluations, which may be useful at future traversal steps or even optimization rounds executed with different traversal parameters. Another extension is to locally re-optimize the worst performers after calibration (see Section \ref{sec:Recalibration}). Another extension is to embed logic that attempts to parallelize calibration of the traversal threads. Finally, a more exotic extension is to apply reinforcement learning to guide traversal, which has been a valuable resource in playing games like Go at superhuman levels \cite{alphago}. 

\begin{figure}
\includegraphics[width=\linewidth]{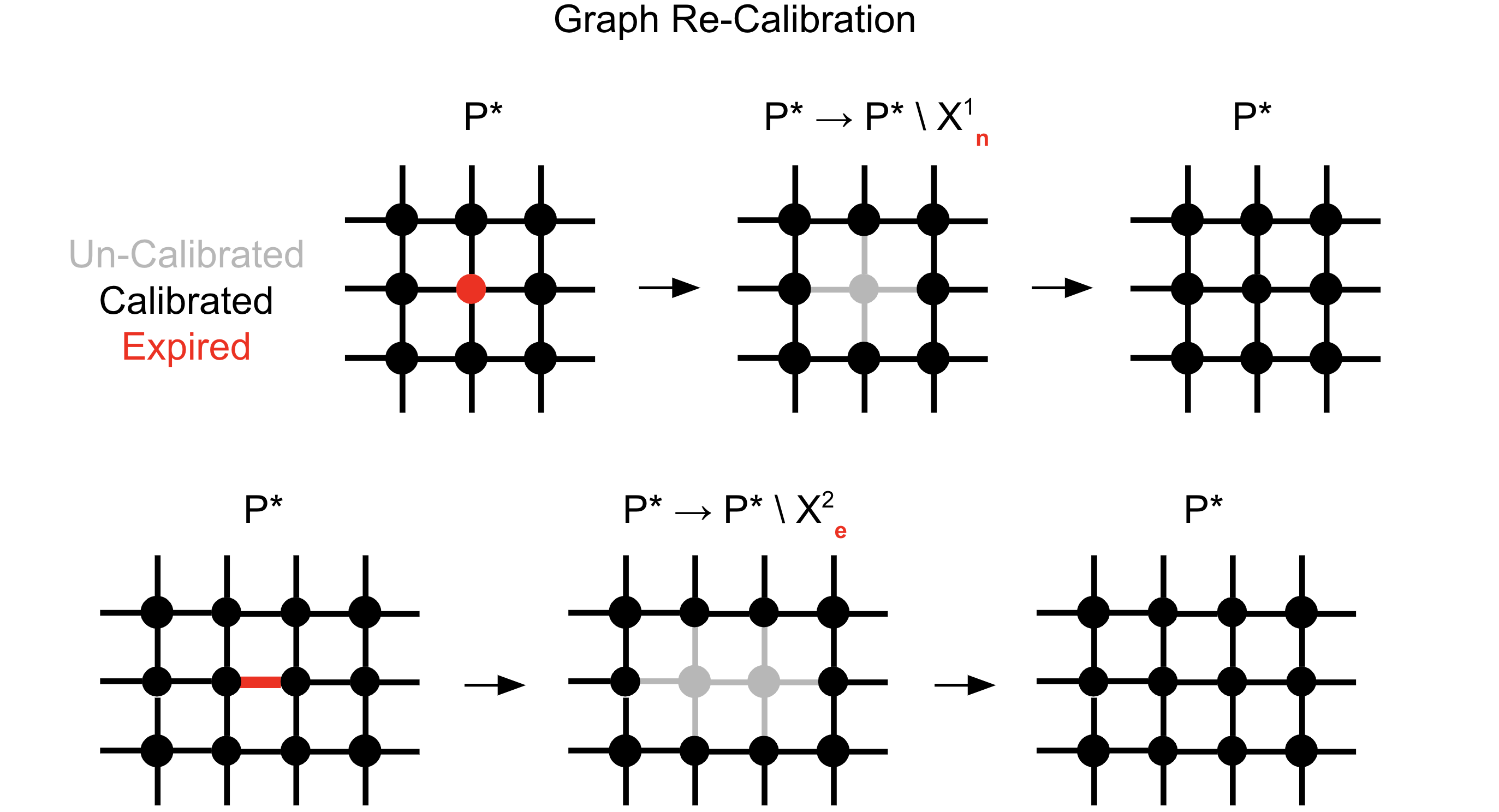}
\caption{\textbf{Graph Re-Calibration.} A graph can be re-calibrated locally to maintain optimal parameters in the presence of system drift.  Here were show how an expired (upper) node and (lower) edge may be re-calibrated. In both cases, we discard nearby graph elements from the calibration status via a connectivity subgraph, and then apply Algorithm \ref{alg:Graph_Calibration}.}
\label{fig:ReCalibration}
\end{figure}

\subsection{\label{sec:Recalibration}Local Re-Calibration}
Optimal calibrated parameters can fluctuate in time. After the fluctuations for some graph element exceed some threshold, as defined by some performance metric, it's respective calibrated parameters are deemed expired. Re-calibrating the full processor in this scenario is undesirable. By design, the Snake is flexible enough to re-calibrate locally while meeting all processor constraints. In particular, when a graph element's calibration has expired, the calibrations in a local neighborhood can be discarded and re-calibrated via Algorithm \ref{alg:Graph_Calibration}, without modification (Figure \ref{fig:ReCalibration}).

\texttt{}

\subsection{\label{sec:Calibration Stitching}Calibration Stitching}
Calibrated regions of a graph that are segmented by un-calibrated regions can be stitched by applying Algorithm \ref{alg:Graph_Calibration}, without modification. In combination with the fact that constraint-disjoint subgraphs may be calibrated in parallel, we can construct advanced calibration procedures. For example, a graph may be intentionally split into multiple constraint-disjoint subgraphs, calibrated in parallel, and finally stitched. Such parallelization strategies will likely be necessary to calibrate large processors.

\section{\label{sec:Outlook}Outlook}
A calibration system must learn optimal control parameters much faster than system drift.  This requirement poses a significant scaling hurdle, since the total number of control parameters typically scales commensurately with processor size. Since the Snake Optimizer can interpolate between optimization complexity and is amenable to both parallelization and local re-calibration, we believe it will serve as a key component in optimizing and operating large scale quantum processors.

\section{\label{sec:Contributions}Contributions}
P.V.K. conceived the Snake Optimizer. P.V.K and J.K. developed ideas for calibration stack integration. J.M.M. and H.N. supported research and development. 

\section{\label{sec:Acknowledgements}Acknowledgements}
Kevin Satzinger, Charles Neill, Chris Quintana, Jimmy Chen, Harry Putterman, and Alexandre Bourassa for implementation contributions and discussions.  

\bibliography{uni.bib}
\end{document}